\def\<{\left\langle}
\def\>{\right\rangle}
\def\td{\mbox{d}}
\def\e{\varepsilon}

\documentclass[twocolumn,showpacs,amsmath,amssymb,aps,floatfix,superscriptaddress]{revtex4}
\usepackage{graphicx,color}
\usepackage{mciteplus}

\begin{document}

\title{The simplest microscopic model of a complex fluid: flow phenomena and constitutive relation}

\author{R. M. L. Evans\footnote{corresponding author}}
\email{mike.evans@physics.org}
\author{Craig A. Hall}
\affiliation{School of Mathematics, University of Leeds, LS2 9JT, United Kingdom}
\author{R. Aditi Simha}
\affiliation{IIT Madras}
\author{Tom Welsh}

\date{August 2016}

\begin{abstract}
It was shown in [PRL {\bf 114}, 138301 (2015)] that a remarkably simple dynamical model exhibits many of the complex flow regimes and non-equilibrium phase transitions characteristic of complex fluids. By removing extraneous detail, this simplest microscopic model of non-Newtonian flow can reveal the universal physics relevant to all complex fluids. Here we present more detailed results and a full derivation of the model's compact mean-field constitutive relation, with great potential scope for insights into universality and tractable mathematics. By enforcing local conservation of angular momentum, the one-dimensional (1D) XY-model (originally used for equilibrium magnetic systems) can be driven into various flow regimes, including simple Newtonian behaviour, shear banding, solid-liquid coexistence and slip-plane motion. The model demonstrates that the phenomenon of shear banding does not rely on details of tensorial stress fields, but can exist in 1D.

\end{abstract}




\maketitle

\section{Introduction}

A detailed and elaborate model of a particular physical systems cannot be used to investigate the physics underlying universal phenomena, because it would not be clear which of the model's features are crucial to the results. Only by removing a particular feature from a model can one ascertain whether it was important. Appreciating this, the pioneers of condensed matter theory investigated universality in equilibrium phase transitions using some archetypal models, including the Ising model and XY model \cite{XY0}.

Experiments and simulations also hint at some universal phenomenology far from equilibrium, in the flow-induced transitions of non-Newtonian fluids. Transitions between Newtonian flow, shear-banding and slip-planes/fractures can be observed in systems as diverse as foams \cite{foamCoex,Debregeas01}, dense colloids \cite{colloidBanding1,colloidBanding2,Haw04}, surfactant solutions \cite{Schmitt94,surfBanding1,surfBanding2,Jones95} and polymers \cite{polyBanding1,polyBanding2,Tapadia03,Pouyan09}. In Ref.~\cite{PRL}, we demonstrated that those same phenomena exist in the one-dimensional (1D) classical XY model (also known as the classical rotor model or $O(2)$ model) when it is given the simplest of angular-momentum-conserving Langevin dynamics. Here we provide more details of the model's steady-state behaviour,  and the full derivation of its remarkably compact and useful constitutive relation.

\begin{figure}
  \includegraphics[width = 5.5cm]{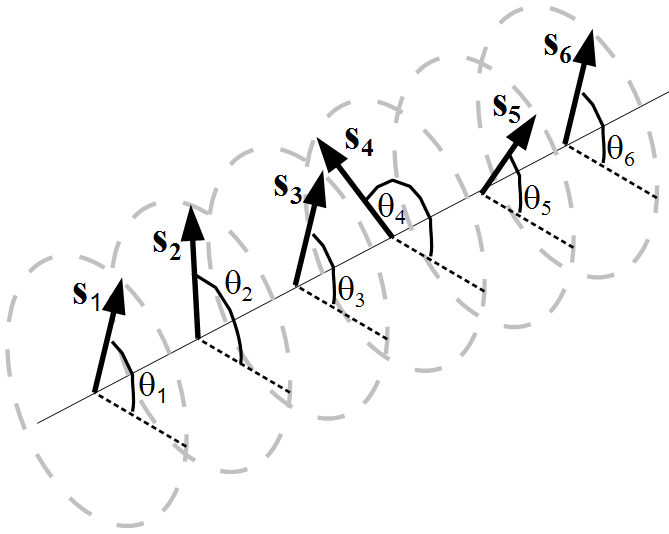}
  \caption{\label{XY}An oblique view of part of the chain of rotors in the one-dimensional classical XY model.}
\end{figure}

As depicted in Fig.~\ref{XY} the 1D XY model consists of a chain of lattice sites, each with one degree of freedom, characterized by the orientation $\theta_j$ of a two-component unit vector ${\bf s}_j=(\cos\theta_j,\sin\theta_j)$, which was originally invented to represent the constant-magnitude magnetisation of an atom with a continuously orientable magnetic moment constrained to lie in the $(x,y)$-plane. 

Each site couples only to its two nearest neighbours, via an interaction potential that favours parallel alignment, so that the model's Hamiltonian (in units of the uniform coupling constant) is given by
\begin{equation}
  H = \sum_{j=1}^N \left[ -{\bf s}_j \cdot {\bf s}_{j-1} + \tfrac{1}{2} \dot{\theta}_j^2 \right]
\end{equation}
where the rotor's moment of inertia (scaling the final, kinetic energy term) has been set to unity without loss of generality. We may impose periodic boundary conditions by defining ${\bf s}_0\equiv{\bf s}_N$, or free boundaries by ${\bf s}_0\equiv{\bf s}_1$, or other possibilities, discussed further below.
As with any 1D system with short-range interactions, the 1D XY model has trivial equilibrium phase behaviour, with a single transition to an ordered state at zero temperature \cite{Mattis84}.

Away from equilibrium, we have found that this simple one-dimensional model has an elaborate steady-state phase diagram with transitions between different flowing phases. By choosing to represent the $(x,y)$-plane perpendicular to the direction of the 1D lattice, we can compare the model's local angular velocity with the rotation flow profile of a fluid in a parallel-plate rheometer \cite{rheometry1,rheometry2}. Of course, our model is not truly three-dimensional like the fluid in a rheometric experiment, but it has many of the same features, including microscopic internal interactions, a macroscopic steady-state flow profile and shear rate, energy exchange, conserved angular momentum and torque propagation.

Equilibrium phase behaviour is independent of the model's dynamics, so long as a Hamiltonian is defined and detailed balance is respected, but non-equilibrium states depend on the specific dynamics. We use the simplest physically motivated dynamics for a thermal system: Langevin dynamics, with all forces (torques) respecting Newton's (rotational) third law of motion. Hence, all forces (conservative, dissipative and stochastic) are equal and opposite on nearest neighbours, and  and angular momentum is conserved. The equations of motion are
\begin{subequations}
\begin{eqnarray}
  \label{EoM}
    \ddot{\theta}_j &=& \tau_j - \tau_{j-1}, \\
  \label{torque}
    \tau_j &=& \sin \Delta\theta_j  + \mu \Delta\dot{\theta}_j + \eta_j(t)
\end{eqnarray}
\end{subequations}
where $\tau_j$ is the torque between rotors $j$ and \mbox{$j+1$}, which have relative angle $\Delta\theta_j = \theta_{j+1}-\theta_j$. The coefficient of friction is $\mu$, and $\eta_j(t)$ are the usual delta-correlated stochastic functions at temperature $T$ with 
\begin{equation}
\label{noise}
	\<\eta_i(t)\eta_j(t')\>=\mbox{$2\mu T\delta(t-t')\delta_{ij}\,$}
\end{equation}
and zero mean. Hamilton's equations are recovered in the conservative case $T=\mu=0$.
Note that no centrifugal or corriolis forces are in effect, because there is no radial component of motion, so the system is independent of absolute angular velocity and respects an angular version of Galilean relativity.

We can nevertheless drive the system into a non-equilibrium steady state by counter-rotating its boundaries to induce shear flow. Thus, non-zero mean work is done by the boundary, but no body-forces are applied, so the Hamiltonian remains that of the XY model. 

We have conducted both numerical and mathematical investigations of the XY model's steady-state behaviour in shear flow. In the next section, we define the numerical protocol, including the physically appropriate boundary conditions for imposing torsional shear flow in a simple manner. The results, including the steady-state phase diagram of the XY model, are presented in section \ref{results}. A theoretical analysis is given in section \ref{theory}, and conclusions and open questions are discussed in section \ref{conclusions}.

\section{Numerical method}

\subsection{Algorithm}

The Dissipative Particle Dynamics (DPD) algorithm \cite{DPD} was used to evolve the state of the system. It shares the stability of the velocity Verlet algorithm, but can be applied to situations with velocity-dependent forces, such as ours, and includes the random thermal forces of Langevin dynamics.

In order to eliminate edge effects, all rotors were treated equally by the simulation algorithm, through the application of periodic boundary conditions. It is perhaps surprising that boundary conditions exist that are periodic but also apply a relative torque across the system. The appropriate condition is an angular version of the Lees-Edwards ``sliding brick" boundary condition \cite{LeesEdwards}, whereby the rotors at opposite ends of the chain are designated as neighbours, but see each other through an angular offset that increases at a constant angular velocity. This can be done without altering Eqs.~\ref{torque} simply by defining
\begin{equation}
  \tau_0 \equiv \tau_N
\end{equation}
and 
\begin{equation}
\label{PERIODICBOUNDRYCONDITION}
 \Delta\theta_N \equiv \theta_0 - \theta_N + N \dot{\gamma}t
\end{equation}
so that
\begin{equation}
  \Delta\dot{\theta}_N = \dot{\theta}_0 -\dot{\theta}_N + N \dot{\gamma}. 
\end{equation}
Hence the $N$-rotor system occupies a periodic space which has a topological twist that increases at a rate $N\dot{\gamma}$. That twist need not be localized between rotors $N$ and $1$, but may be distributed across the system, while the boundary rotors may choose their velocities such that their interaction torque $\tau_N$ fluctuates in a statistically identical way to the other inter-rotor torques. Thus all rotors are governed by the same equations of motion, while the system as a whole experiences a shear rate (angular velocity difference per rotor) of $\dot{\gamma}$.

The distribution of the random torques $\eta_j$ was given a Gaussian form
with variance 
\begin{equation}
  \label{RANDOM:SD}
  \sigma^2 =\frac{2\mu\,T}{dt}
\end{equation}
and zero mean and a discrete time step $dt$.

\subsection{Calibration}

The time step $dt$ was chosen adaptively to ensure that the interaction potential was explored in sufficient detail, irrespective of rotor speed. On each time step, the maximum increment in angular difference between any interacting pair of rotors was limited to a maximum magnitude $\Delta\theta_{\rm max}$. Thus, no pair of rotors could pass over the potential barrier without feeling the appropriate torque. If a pair of interacting rotors had a sufficiently high relative angular velocity that their relative angular increment exceeded the specified limit, all data for that time-step was discarded and the step was repeated with a reduced time interval $dt$. To maintain simulation speed, the time-step for the next iteration was increased whenever the maximum relative angular increment between all neighbours was found to be substantially below the threshold.
For the majority of states, results were found to be reliable for a threshold as high as 1~radian. This becomes less surprising when one considers that the vast majority of relative angular increments within the system are much smaller than the threshold value. Nevertheless, we chose never to set $\Delta\theta_{\rm max}$ above $0.2$, and used lower values in some cases (often $\Delta\theta_{\rm max}=0.1$).

To ensure that no discretization artefacts, start-up transients or finite-size effects were present in the final results, trial simulations were run in a variety of different regions of the $(T,\mu,\dot{\gamma})$ parameter space, and repeated for different values of $\Delta\theta_{\rm max}$ and system size $N$, and different initial conditions.
A system of $N=256$ rotors was found to be sufficiently large to eliminate finite-size effects from all states except for those with very widely separated slip-planes (see Results, section \ref{results}). Hence a safe value of $N=512$ was used for most simulations, occasionally increasing to 1024 or 2048 for extreme cases. 

In experiments and simulations, both in and out of equilibrium, it is never straightforward to determine the decay time of all start-up transients in order categorically to establish a steady state. For instance, equilibrium investigations can be misled by long-lived metastable states that are impossible to distinguish from the true steady state except by infinite patience. (Without evidence of the existence of graphite, diamond would be mistaken for the ground-state of carbon.) We must therefore accept that our study of the XY model's steady states cannot be definitive. Subject to that proviso, we judge that initial transients have decayed when simulations with very different initial conditions converge to a statistically similar state of motion. The initial conditions tried included a state of uniform shear rate ($\Delta\dot{\theta}_i=\dot{\gamma}\;\forall~i$) and states in which plateaux of uniform velocity (where $\dot{\theta}_i=\mbox{constant}$) are separated by sharp discontinuities (``slip planes"). The convergence time varies greatly with the values of the parameters ($T,\mu,\dot{\gamma}$), becoming very large close to phase transitions. In most cases the initially uniform system arrived most quickly at its final steady state.

Having established our protocols, we performed simulations at a large number of points in parameter space in order to chart the system's non-equilibrium steady-state phase behaviour.

\section{Results}
\label{results}

\begin{figure}
  (a) \hfill \\ 
  \includegraphics[width=4.2cm]{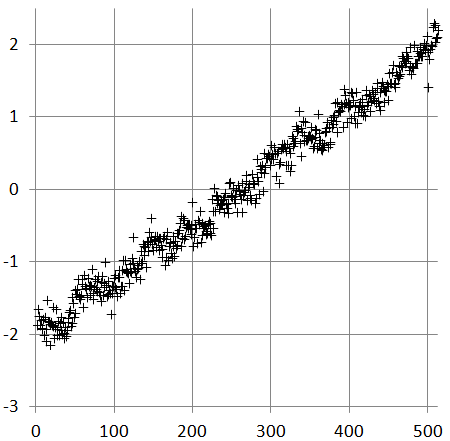}
  \includegraphics[width=4.2cm]{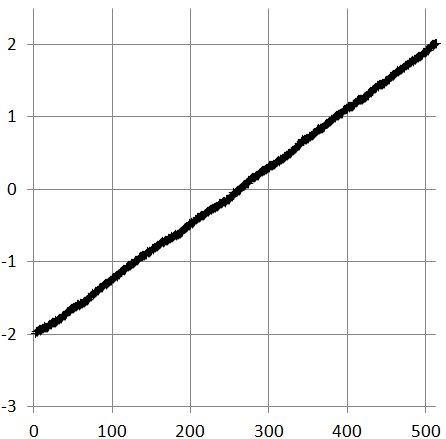} \\
  (b) \hfill \\ 
  \includegraphics[width = 4.2cm]{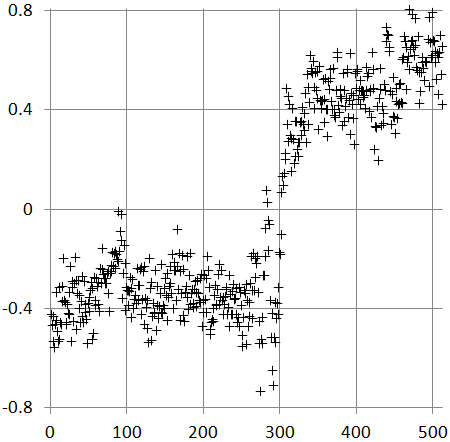}
  \includegraphics[width = 4.2cm]{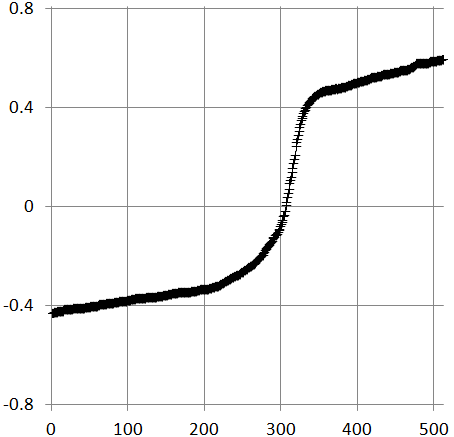} \\
  (c) \hfill \\ 
  \includegraphics[width = 4.2cm]{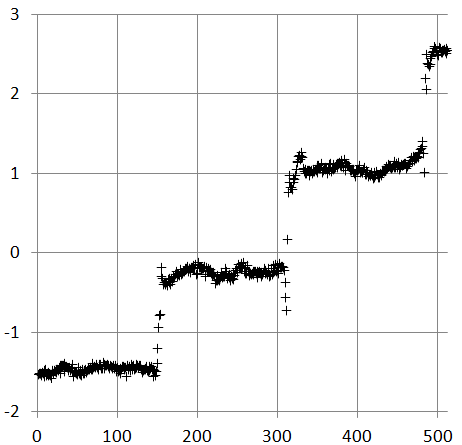}
  \includegraphics[width = 4.2cm]{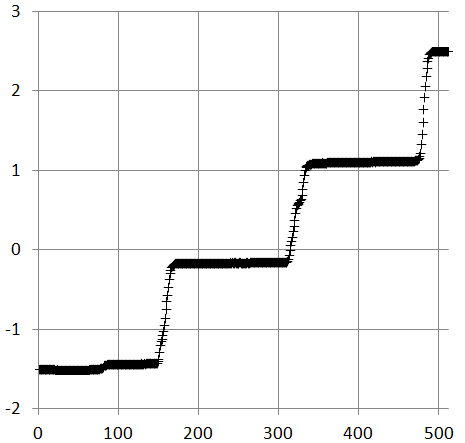} \\
  (d) \hfill \\ 
  \includegraphics[width = 4.2cm]{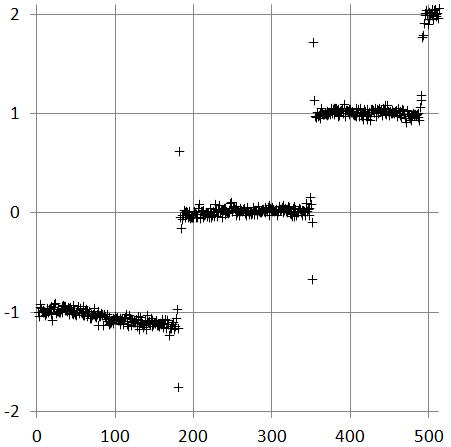}
  \includegraphics[width = 4.2cm]{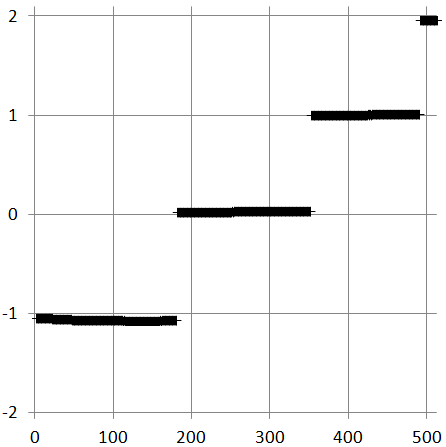} \\
  \caption{\label{examples}Angular velocity as a function of position along the one-dimensional chain of 512 rotors for various parameter values, typifying the four qualitatively distinct steady states. Left-hand column: instantaneous snap-shots of angular velocity. Right-hand column: averaged over several rotation-times ($\sim\dot{\gamma}^{-1}$). 
  (a) Uniform shear flow at $(T,\mu,\dot\gamma)=(0.02, 10, 0.007813)$;
  (b) shear banding at $(T,\mu,\dot\gamma)=(0.006, 1, 0.002)$.
  (c) a solid-fluid coexistence at $(T,\mu,\dot\gamma)=(0.001, 1, 0.0078125)$;
  (d) slip planes at $(T,\mu,\dot\gamma)=(0.001, 0.5, 0.00585938)$;
  }
\end{figure}

Four distinct types of flow behaviour where found, separated by phase transitions or narrow cross-over regions: a state of uniform time-averaged shear flow, a shear-banding state, a coexistence between solid and fluid regions, and a state of solid regions separated by localized slip planes. Examples of each are shown in Fig.~\ref{examples}, in steady states that were found to be well established after $t=4500$. Animations of some of the steady states are available online \cite{animations}.

On increasing temperature or friction coefficient or shear rate, the general trend is to progress from the slip-plane regime, through solid-fluid coexistence and then the shear-banding regime, to a state of statistically uniform shear flow. However, for some parameter values, one or both of the two intermediate states (solid-fluid coexistence and shear-banding) are bypassed.

While a wide variety of relative angles, velocities and accelerations exist at any instant within the uniform phase, it is {\em statistically} uniform, in the sense that 
$\<\td\Delta\theta_j/\td t\>$ 
is independent of $j$, where $\<\ldots\>$ indicates a time-average. The uniform limit at large parameter values is easily explained; all structure due to interactions is washed out when either thermal energy or shear-induced kinetic energy of relative motion swamps the scale of the interaction potential, or when frictional forces dominate over the potential gradient. This argument, for high shear rate, will be made more rigorous in section~\ref{bounds}, and a full theoretical investigation of the phase behaviour will be presented in section \ref{theory}.

\subsection{Slip planes}

\begin{figure}[h]
  (a) \hfill \\ 
  \includegraphics[width = 5.5cm]{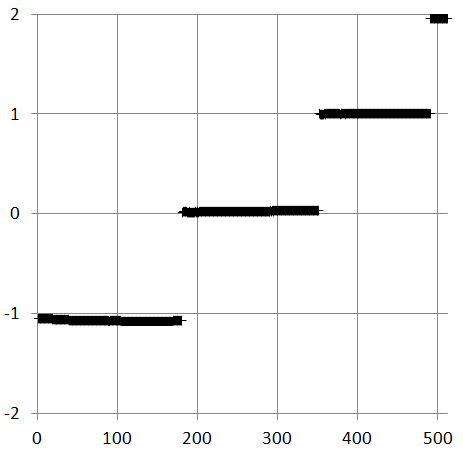} \\
  (b) \hfill \\ 
  \includegraphics[width = 5.5cm]{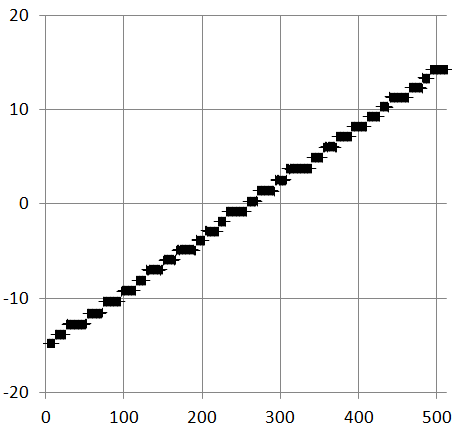} \\
  (c) \hfill \\ 
  \includegraphics[width = 5.5cm]{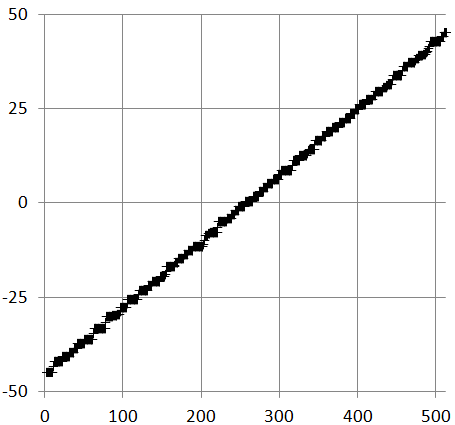} \\
  \caption{\label{slipPlaneProgression} A selection of time-averaged angular velocity profiles (as a function of position) in the slip-plane phase. As overall shear rate increases at fixed $T=0.001$ and $\mu=0.5$, the discontinuities in velocity remain approximately constant in magnitude but increase in number. (a) $\dot\gamma=0.005859$; (b) $\dot\gamma=0.05859$; (c) $\dot\gamma=0.175781$.
  }
\end{figure}

The existence of slip planes takes more consideration. Some velocity profiles in the slip-plane state are shown in Fig.~\ref{slipPlaneProgression} for parameter values $T=0.001$ and $\mu=0.5$ and various shear rates $\dot\gamma$.

In the slip-plane regime, the majority of rotors have a time-averaged angular velocity equal to that of their neighbours. So the inter-rotor angle $\Delta\theta$ exhibits only small-amplitude excursions about a constant value, within the potential well. These rotor-pairs are ``locked" in the terminology of B\"{u}ttiker et al.~\ref{Buttiker}. Hence, most of the system behaves as an elastic solid, while all of the shear flux is concentrated in a small population of isolated inter-rotor gaps (the slip planes), where the relative angle is ``running" in the terminology of B\"{u}ttiker et al.~\cite{Buttiker}. Crucially, the position of each slip plane remains fixed for a long duration, far exceeding its internal rotation time $2\pi/\Delta\dot{\theta}_i$,  so that the pair of rotors straddling it execute many full turns of relative motion before the slip plane stochastically wanders to a new site. Indeed in the majority of cases studied, after initial transients, slip planes remained fixed for the entire simulation, whilst traversing their potential barrier many thousands of times.

\begin{figure}[h]
  \includegraphics[width = 7.5cm]{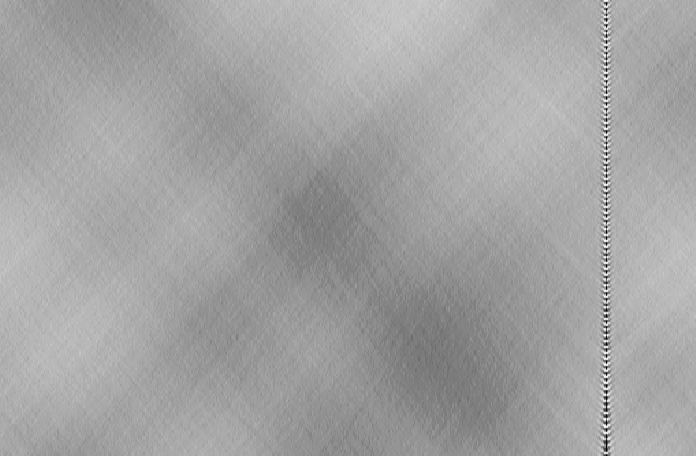}
  \caption{\label{space-time-SLIP}The entire system, from rotor number 1 to 512, is represented across the width of the picture. Time proceeds vertically, spanning an interval of 496 time units in this image. Grey scale representing inter-rotor potential, from $U=-1$ (white) to $U=1$ (black). The steady state shown (following an initialization time in excess of 4500 time units) is at parameter values $(T, \mu, \dot{\gamma})=(10^{-4}, 0.5, 0.001896)$, for which the system exhibits a single slip plane.}
\end{figure}

The stability of each solid region or ``plateau" within the steady-state requires the overall time-averaged torque within the system (applied by the boundary condition) to be smaller than the steepest gradient in the interaction potential (unity), which is the threshold torque at which an elastic solid region would yield. (This follows from time-averaging Eq.~\ref{torque} given that, within this region, $\Delta\theta$ remains bounded, by definition of ``solid", so that its averaged time-derivative vanishes.) Although the time-averaged torque is uniform at steady state (from time-averaging Eq.~\ref{EoM}), and smaller than the yield threshold, it is nonetheless sufficient to maintain flow within the slip planes, where the potential barrier is repeatedly overcome by virtue of the rotors' momenta. The traversing of the periodic potential creates a periodic variation in torque at the slip plane, which propagates outwards as damped torsion waves in the elastic solid regions.

This motion can be seen in on-line animations \cite{animations}
and is also visible in Fig.~\ref{space-time-SLIP}, which is a space-time diagram of the system's motion. A grey scale represents the value of the inter-rotor potential $U(\Delta\theta)=-\cos\Delta\theta$ at each of the 512 sites along the chain (set out horizontally in the figure). Time increases vertically in the figure.  

In Fig.~\ref{space-time-SLIP}, all of the shear is concentrated in a single slip plane between two rotors, at a location (selected by spontaneous symmetry breaking in the system) near the right-hand end of the chain. The slip plane is clearly distinguished in the space-time diagram by the repeated set of black marks where the maximum of the potential is repeatedly traversed. High-frequency waves (visible as short slanting lines) propagate outwards only a short distance to either side of the slip plane, before decaying away. Meanwhile very long wavelength torsional waves of lower amplitude travel across the entire system length, visible as long, faint slanting marks in Fig.~\ref{space-time-SLIP}. The gradient of the marks indicates a wave speed of approximately $0.71$ in this case.

\subsection{Solid-fluid coexistence}

\begin{figure}
\begin{center}
  \includegraphics[width = 7.5cm]{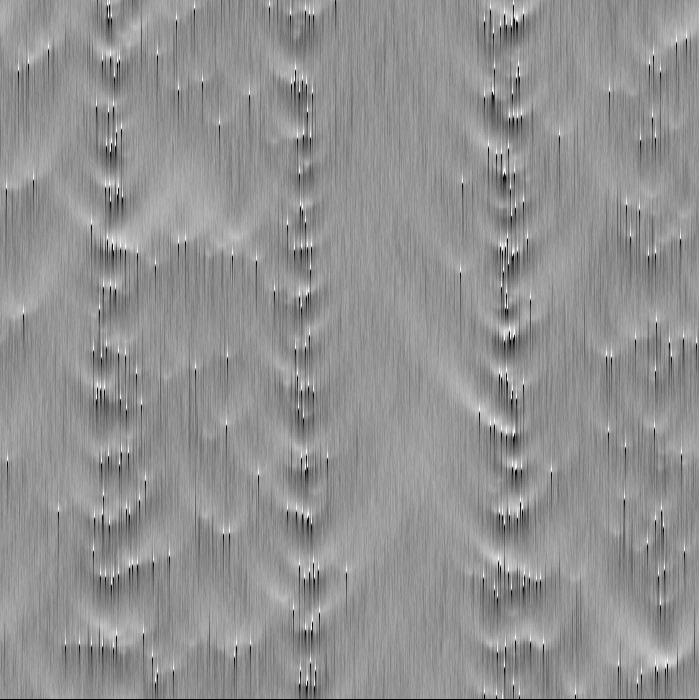}
  \caption{\label{space-time-COEX}Space-time diagram representing inter-rotor potential (from $U=-1$ (white) to $U=1$ (black)) as a function of position (horizonal axis) and time (increasing vertically)  (as Fig.~\protect\ref{space-time-SLIP}) and parameter values  $(T, \mu, \dot{\gamma})=(0.0005,5,0.002)$, for which the steady state is a coexistence between fluid and solid regions. The vertical axis covers a duration of 1500 time units, beginning 4500 time units after initialization. The system contains $N=1024$ rotors, arranged along the horizontal axis.}
\end{center}
\end{figure}

Whereas the space-time diagram for slip planes (Fig.~\ref{space-time-SLIP}) shows linear propagation of small-amplitude waves within the solid regions, large-amplitude non-linear waves of yielding events propagate through the fluid region of a solid-fluid coexistence. This can be seen in Fig.~\ref{space-time-COEX}, the space-time diagram for a steady state of solid-fluid coexistence, where the black marks indicate events where the maximum of the potential is overcome. 

It is clear, in Fig.~\ref{space-time-COEX}, that each yielding event locally displaces stress sufficient to trigger another yielding event nearby, leading to a propagating sequence of rotor-pairs crossing their potential barrier within a finite region of the system. The stress excitations from these events do not propagate across the whole system. The (uniform) time-averaged torque remains just below the yield-point value of unity (the maximum of $\sin\Delta\theta$ in Eq.~\ref{torque}), so a portion of the system is able to exist in a solid state. In fact, due to the finite temperature, some relatively rare yield events are visible in the ``solid" region also, hence, it is not strictly solid, but a high-viscosity fluid, so that this coexistence may not be qualitatively distinct from a shear-banding state.

It is surprising that this one-dimensional system has sufficient complexity to self-organise into a stable coexisting state. Some light is shed on the mechanisms responsible for this phenomenon by the mean-field theory in section \ref{EMT}.

\subsection{Shear banding and critical states}
\label{bandingAndCritical}

In the shear-banding state, the X-Y model self-organises into macroscopic regions with different effective viscosities.
As in solid-fluid coexistence (which we have seen is a special case of shear banding with a large ratio of effective viscosities), the stress in the shear-banding state is only a little below its maximum (yield-point) value of unity so that a small stress perturbation can trigger a large-amplitude rearrangement.

\begin{figure}[h]
\begin{center}
  \includegraphics[width = 7.5cm]{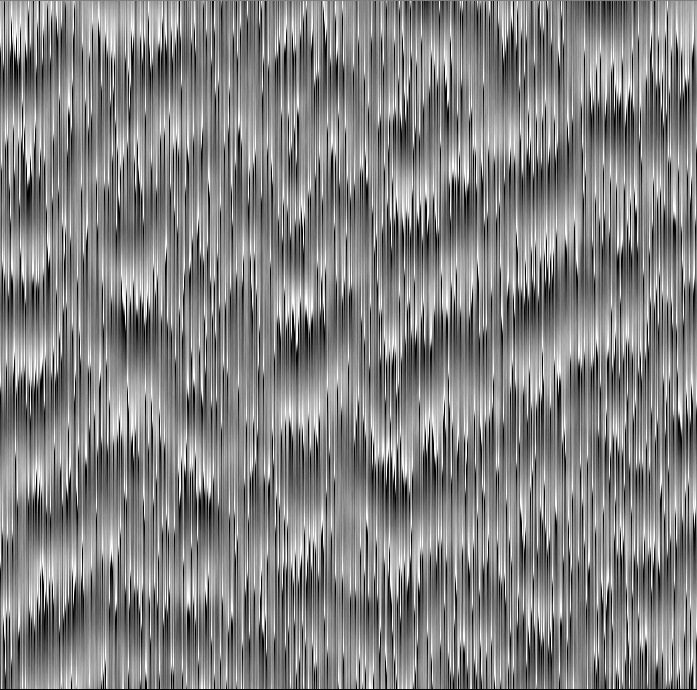}
  \caption{\label{space-time-CRIT}The entire system is represented across the width of the picture, from rotor number 1 to 1024, following an initialization time in excess of 65000 time units. Time proceeds vertically, spanning an interval of 750 time units in this image. Grey scale representing inter-rotor potential, from $U=-1$ (white) to $U=1$ (black). The steady state shown is at parameter values $(T, \mu, \dot{\gamma})=(10^{-3}, 10, 0.017578125)$, for which the system exhibits a statistically uniform phase but contains correlations of long range in space-time, due to the proximity of the continuous transition to banding.}
\end{center}
\end{figure}

It appears that a continuous transition separates the shear-banded and uniform states. However, we have not established whether this is an isolated critical point, a line of continuous transitions, or a critical point that terminates a line of first-order phase transitions. The space-time diagram in  Fig.~\ref{space-time-CRIT} is for a system close to criticality, at the transition between uniform and shear-banded states. 

A number of critical phenomena are apparent here. One characteristic feature of a nearby critical point is a large correlation length, indicating highly collective/cooperative motion. Although diverging correlation lengths are not possible in one-dimensional, locally-interacting systems at equilibrium, our non-equilibrium steady state exhibits this feature. It can only be recognized by observing the space-time domain, as in Fig.~\ref{space-time-CRIT}, where regions of correlated configurations (in which $\Delta\theta\approx\pi\;\Rightarrow\;U\approx1$) extend across large distances, but not at constant time.

\begin{figure}
  \includegraphics[width = 8.3cm]{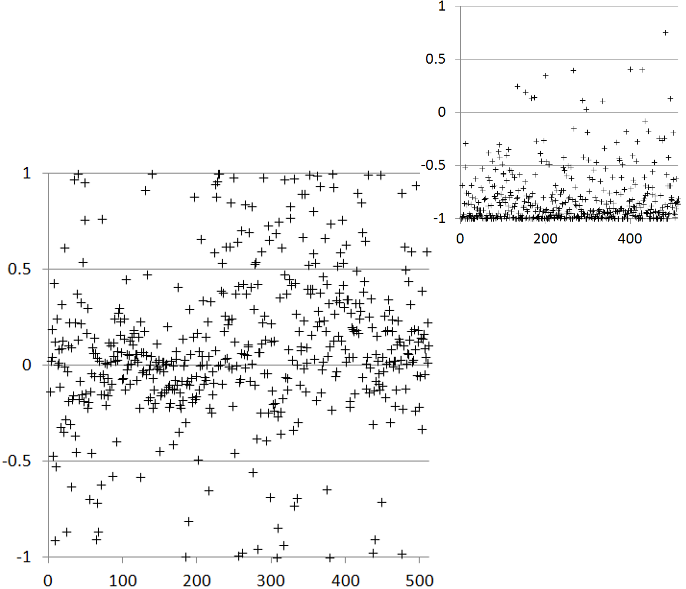}
  \caption{\label{critical-potential}Value of the interaction potential $U=-\cos(\Delta\theta)$ as a function of position (inter-rotor gap index $j$ running from 1 to 512, showing half of the system), for the near-critical uniform fluid at  $(T, \mu, \dot{\gamma})=(10^{-3}, 10, 0.017578125)$, as shown in Fig.~\ref{space-time-CRIT}b. {\bf Inset:} For comparison, the same plot for an equilibrium system at  $(T, \mu, \dot{\gamma})=(0.3, 1, 0)$ in which the potential is occupied according to Bolzmann's law, with the lowest value most likely.}
\end{figure}

Secondly, the very slow decay of initial transients is typical of critical slowing-down. A third critical phenomenon is the divergence of response functions (discussed more in section \ref{PDStressEnergy} and Fig.~\ref{stress-strain}b), due to the fact that the system appears to be poised in a highly susceptible state.
Fig.~\ref{critical-potential} shows the value of the interaction potential $U=-\cos(\Delta\theta)$ for each rotor pair. The yield point on this potential, at which a stable elastically-bound interaction fails because $U''\to0$, is at $\Delta\theta=\pi/2$, where $U=0$, i.e. halfway up the potential barrier. Note that the system forms a distribution that is clustered around this yield point. By comparison, the inset of Fig.~\ref{critical-potential} shows the values of $U$ in a system at equilibrium ($\dot{\gamma}=0$), which are Boltzmann-distributed close to the potential minimum at $U=-1$.

\subsection{\label{PDStressEnergy}Phase diagrams and measurements of stress and energy }

\begin{figure}
\begin{center}
  (a) \hfill \\ \includegraphics[width = 7cm]{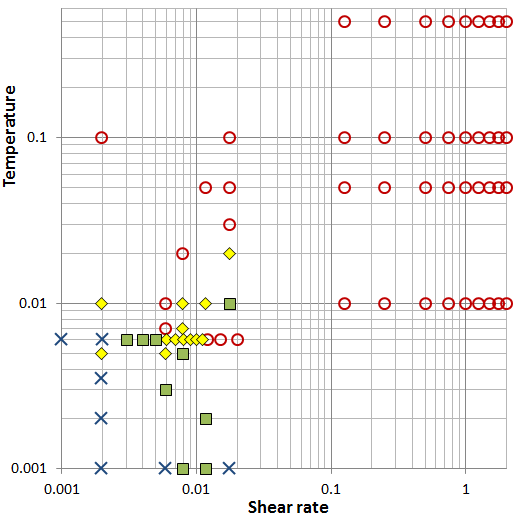} \\
  (b) \hfill \\ \includegraphics[width = 7cm]{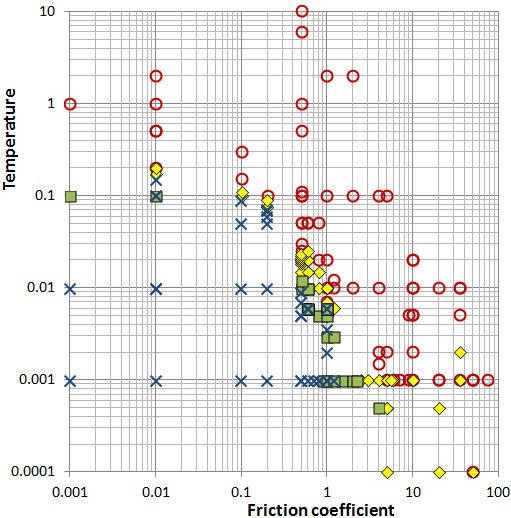}  \\
  (c) \hfill \\ \includegraphics[width = 7cm]{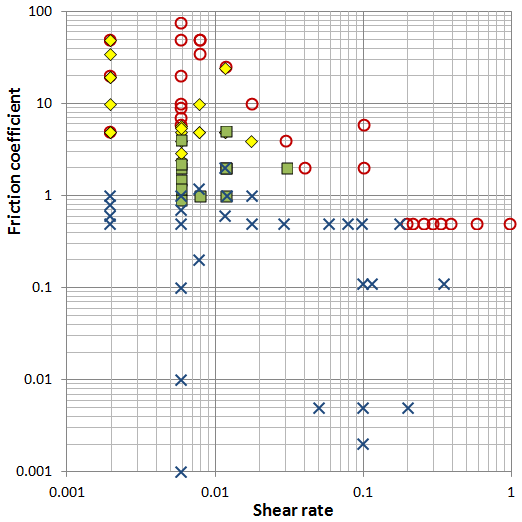} \\
  \caption{\label{phase_diagrams}(Colour on-line) Steady-state phase diagrams for the angular-momentum-conserving XY model, shown on representative slices through the parameter space. Steady states are represented by red open circles (uniform phase); yellow filled diamonds (shear banding); green filled squares (solid-fluid coexistence); blue crosses (slip-plane phase). Data are shown for simulations at parameter values:  (a) $\mu=1.0$; (b) $10^{-3}\leq\dot\gamma\leq~10^{-2}$; (c) $10^{-4}\leq~T\leq~10^{-3}$.}
\end{center}
\end{figure}

Representative results are shown on some mutually orthogonal slices through the parameter space in Fig.~\ref{phase_diagrams}. In most cases, the qualitative type of flow behaviour was identified, from time-averaged velocity profiles such as those in Fig.~\ref{examples}. The identification was unambiguous in most cases, but the distinction between shear-banding states and uniform or solid-liquid-coexisting states was sometimes unclear, particularly near criticality. 

\begin{figure}[h]
\begin{center}
  \includegraphics[width = 7.0cm]{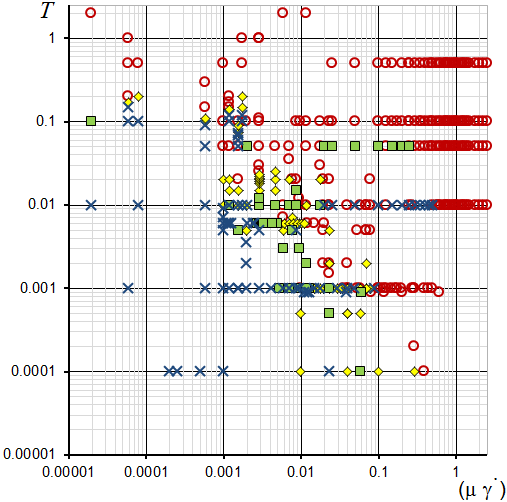}  \\
  \caption{\label{collapse}(Colour on-line) All of our simulated steady-state phase data, on logarithmic axes $T$ versus $\mu\dot{\gamma}$. The phases are represented by: blue crosses (slip planes); green-filled squares (solid-fluid coexistence); yellow-filled diamonds (shear banding); red circles (uniform).}
\end{center}
\end{figure}

It is informative to plot data from all of our steady-state simulations on axes $T$ versus $\mu\dot\gamma$, allowing us to present all of our simulated phase data in one figure, while Fig.~\ref{phase_diagrams} contains only a subset. This is shown in Fig.~\ref{collapse}.
The simulations were performed at values of $\mu$ ranging from $10^{-3}$ to $10^2$ and $\dot{\gamma}$ from $10^{-3}$ to $7.0$. Clearly, there is not a full data collapse, but some structure is visible in Fig.~\ref{collapse}
--- in particular, the uniform region for values greater than $1$ on either axis, and general trends along the major and minor diagonals.
Further understanding, and some justification for a partial data collapse on axes of $T$ versus $\mu\dot\gamma$ are provided by the upper bounds in section \ref{bounds} and the approximate scaling \mbox{analysis} in \mbox{appendix \ref{scaling}}.

\begin{figure}
\begin{center}
  \includegraphics[width = 8.7cm]{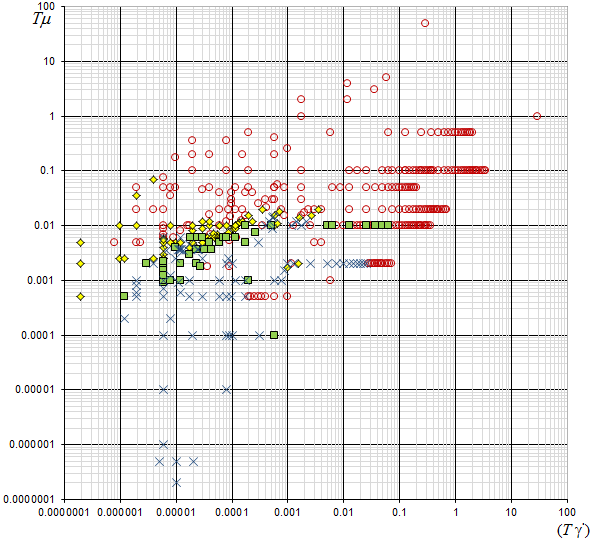}  \\
  \caption{\label{inexplicable}(Colour on-line) All of our simulated steady-state phase data, on logarithmic axes $T\mu$ versus $T\dot{\gamma}$. As in Fig.~\protect\ref{collapse}, the phases are represented by: blue crosses (slip planes); green-filled squares (solid-fluid coexistence); yellow-filled diamonds (shear banding); red circles (uniform).}
\end{center}
\end{figure}

Empirically, we find a marginally better collapse of the data into fairly clearly-defined phase regions on axes of $(T\dot\gamma)$ and $(T\mu)$, as shown in Fig.~\ref{inexplicable}, although we have no explanation for this.

\begin{figure}
  (a) \\  \includegraphics[width = 7cm]{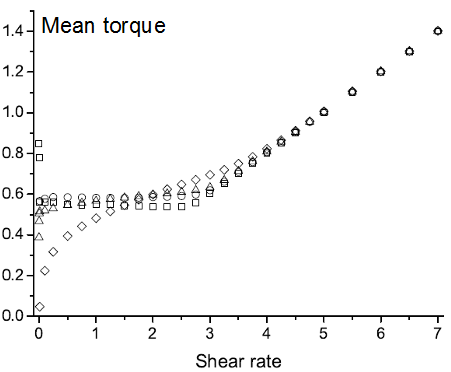}\\
  (b) \\  \includegraphics[width = 7cm]{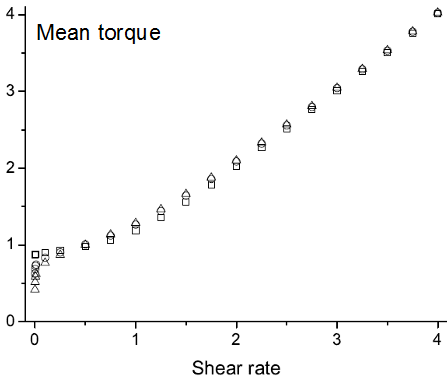}
  \caption{\label{stress-strain}Measurements of the steady-state mean torque $\overline\tau$ as a function of strain rate (shear rate $\dot\gamma$) from simulations at (a) $\mu=0.2$ and (b) $\mu=1.0$. Symbols denote different temperatures: $T=0.01$ (squares); $T=0.05$ (circles); $T=0.1$ (triangles); $T=0.5$ (diamonds). The discontinuous gradient in (a) is reminiscent of crossing a first-order phase transition, while (b) resembles an approach to a critical point.}
\end{figure}

Measurements of the mean torque in the system, $\overline\tau$, as a function of shear rate $\dot\gamma$ are shown in Fig.~\ref{stress-strain}. These are reminiscent of stress-versus-strain-rate measurements of non-Newtonian fluids measured in rheometric studies of their constitutive relations (e.g.~Ref.~\cite{Schmitt94}). A simple Newtonian fluid, on the other hand, would produce a straight line through the origin with gradient equal to its viscosity.

\begin{figure}[h]
    \includegraphics[width = 8.5cm]{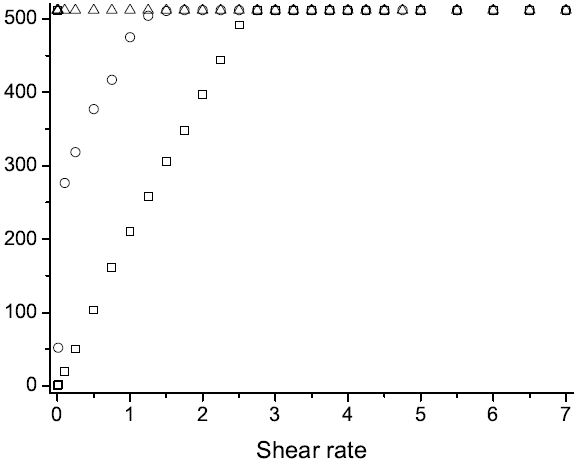} \\
	\caption{\label{slipPlaneCount} The number of running (fluid) pairs (as opposed to solidly locked pairs) measured in simulation of size $N=512$ for $\mu = 0.2$, as a function of shear rate.	Symbols denote different temperatures $T$: squares 0.01; circles 0.05; triangles 0.1.}
\end{figure}

At $\mu=0.2$ and $T=0.01$, the systems crosses a phase transition directly from the uniform state at high shear rate $\dot\gamma$ to the slip-plane state at low shear rate. Notice that, in the slip-plane state, the mean torque remains very constant, independent of $\dot\gamma$. It achieves this by varying the number of slip planes (running pairs) as shear rate varies, as shown in the measurements in Fig.~\ref{slipPlaneCount}, and also in the velocity profiles of Fig.~\ref{slipPlaneProgression}.
At $\mu=0.2$ and $T=0.05$, the data in Fig.~\ref{stress-strain}a cross a phase transition from a uniform phase at high shear rate to a solid-fluid coexistence at low shear rate, while, at $T=0.5$ (and possibly also at $T=0.1$), the system remains uniform.

For the higher value of the friction coefficient $\mu=1$ shown in Fig.~\ref{stress-strain}b, the system remains in the uniform phase for the temperatures and shear rates shown. Note that the axes are both linear, so very low shear rates are not interrogated by this graph. Despite remaining statistically uniform, the system exhibits large-scale fluctuations, particularly for the data at $T=0.01$, as it is close to a continuous phase transition, which we believe to be the same critical point that influenced the data in Fig.~\ref{space-time-CRIT}, discussed in section \ref{bandingAndCritical}. The precise location of this critical point (or line) is difficult to establish, as its critical phenomena are observed across a large region of the parameter space. 

Close to criticality,  response functions diverge. In this case, we can identify the response of the strain rate to changes in stress,
$\left(\partial\dot{\gamma}/\partial\overline{\tau}\right)_{T,\mu}$ as the reciprocal of the gradient in Fig.~\ref{stress-strain}b, which diverges as a continuous transition (possibly to a shear-banded state) is approached from above.

\begin{figure}[h]
  (a) \\  \includegraphics[width = 7.5cm]{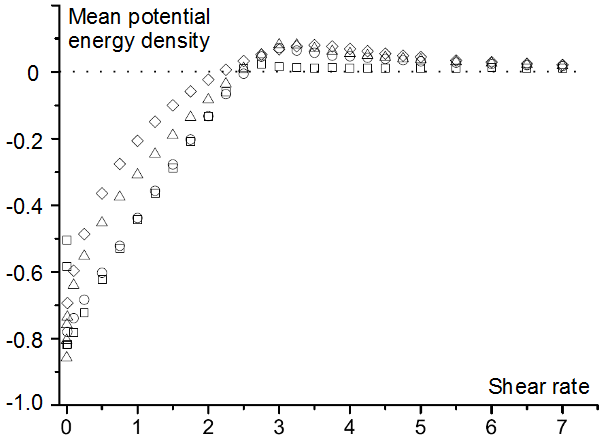}\\
  (b) \\  \includegraphics[width = 7.5cm]{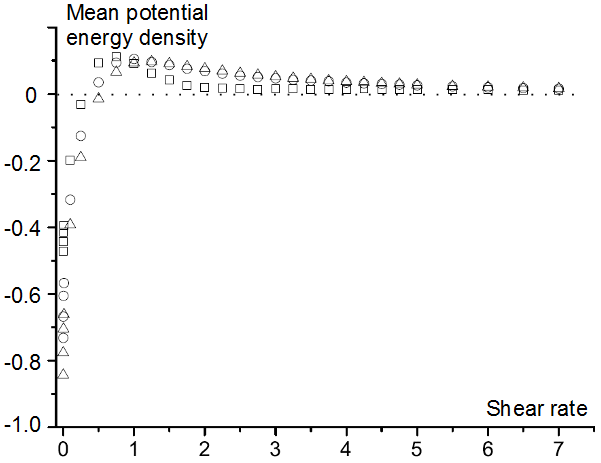}
	\caption{\label{energyFig} Measurements of the mean potential energy density as a function of shear rate, at the same parameter values as in Fig.~\protect\ref{stress-strain}:  (a) $\mu=0.2$ and (b) $\mu=1.0$ Symbols denote different temperatures: $T=0.01$ (squares); $T=0.05$ (circles); $T=0.1$ (triangles); $T=0.5$ (diamonds).}
\end{figure}

 Figure \ref{energyFig} shows the mean potential energy density $\overline{u}$ as a function of shear rate. We define $\overline{u}$ as the value of $-\cos\Delta\theta_j$ averaged with respect to position $j$ and time. Potential energy density is a more suitable thermodynamic quantity to study than the total internal energy density, because the kinetic energy of a continuously sheared system is non-extensive. Features, due to phase transitions being crossed or narrowly avoided, are again visible in these data.

\section{Theoretical analysis}
\label{theory}

\subsection{Steady-state conditions}

Equation.~(\ref{EoM}) yields an equation of motion for the relative angle $\Delta\theta_j = \theta_{j+1}-\theta_j$ between neighbouring rotors,
\begin{equation}
\label{gapEoM}
  \frac{\td^2\Delta\theta_j}{\td t^2} = \tau_{j+1}+\tau_{j-1}-2\tau_j
\end{equation}
which, together with Eq.~(\ref{torque}), forms a closed set of equations in the relative angles $\Delta\theta_j$ only, independent of any absolute angular positions $\theta_j$, as would be hoped for this rotationally symmetric system.

Time-averaging both sides of Eq.~(\ref{gapEoM}), and imposing the steady-state condition that, while rotors may have a finite time-averaged angular velocity, their time-averaged angular acceleration must vanish, yields a balance in mean torque-differences $\<\tau_{j+1}-\tau_j\>=\<\tau_j-\tau_{j-1}\>=\mbox{constant}$. Applying the periodic boundary condition, that the torques are equal at opposite ends of the chain, sets the constant to zero, so that the time-averaged torque is uniform (independent of index $j$) throughout the steady-state system,
\begin{equation}
\label{uniformTorque}
  \<\tau_j\>=\overline{\tau}\quad\forall\; j.
\end{equation}
This resembles the uniform-stress condition for steady shear flow of a fluid.

Let us define the {\em local shear rate} $s_j$ to be that part of the neighbours' relative velocity that is persistent rather than fluctuating:
\begin{equation}
	s_j \equiv \left\langle\Delta\dot{\theta}_j\right\rangle
\end{equation}
which is time-independent, due to the temporal averaging $\langle\ldots\rangle$.
Note that the {\em local} shear rate $s_j$ should not be confused with the {\em global} shear rate $\dot{\gamma}$. The two are equal in the uniform phase. In other phases, $\dot{\gamma}$ is equal to the positional average (with respect to $j$) of $s_j$.

Now time-averaging the torque Eq.~(\ref{torque}) and applying Eq.~(\ref{uniformTorque}) and the fact that the noise has zero mean yields a relationship between the local shear rate and the overall mean torque,
\begin{equation}
\label{stressStrain}
  \overline{\tau} = \mu s_j + \<\sin\Delta\theta_j\>.
\end{equation}
Equation~(\ref{stressStrain}) can be regarded as a rheological stress-strain-rate relation. The final term in Eq.~(\ref{stressStrain}) would vanish for constant relative rotational motion, yielding a Newtonian fluid of viscosity $\mu$, the constant of proportionality between ``stress" (i.e.~torque in this rotational case) and strain rate. However, quasi-periodic velocity fluctuations synchronised with (i.e.~having a frequency component equal to) the local shear rate lead to a non-zero average of $\sin\Delta\theta_j$ due to a non-uniform angular distribution of $\Delta\theta_j$, and hence to non-Newtonian rheology.

\subsection{Upper bounds}
\label{bounds}

Clearly the magnitude of the final term in Eq.~(\ref{stressStrain}) cannot exceed unity. Furthermore, it is reasonable to assume that all terms in Eq.~(\ref{stressStrain}) are positive for a positive overall shear rate $\dot{\gamma}>0$, since the second law of thermodynamics forbids a negative response of torque to imposed shear rate, and local shear rate is expected to share the direction of the global shear rate, as is the conservative part of the torque. Hence $0\leq\<\sin\Delta\theta_j\><1$. 

In any part of the chain that is behaving as a solid, by definition, the local shear rate (the first term on the RHS of Eq.~(\ref{stressStrain})) vanishes, yielding a condition on the overall torque $\overline{\tau}<1$ for the existence of any solid regions. Beyond this threshold, no bound structures, even as small as a single pair of rotors, can survive. Hence we anticipate a uniform state when the condition is violated.

The fluid regions (portions of the chain in which the local shear rate is non-zero) also put conditions on $\overline{\tau}$. Since the positionally-averaged local shear rate is constrained to equal $\dot{\gamma}$, the maximum (with respect to $j$) local shear rate cannot be less than $\dot{\gamma}$. Hence, at that fastest-shearing location, Eq.~(\ref{stressStrain}) gives $\overline{\tau}>\mu\dot{\gamma}$. Combining this with the other inequality on $\overline{\tau}$, and recalling that $\overline{\tau}$ is independent of position $j$, implies that the steady state is uniform if $\mu\dot{\gamma}>1$. 

This upper bound on non-uniform states is consistent with the data in Fig.~\ref{collapse}.
The highest value of $\mu\dot{\gamma}$ for which non-uniform behaviour was identified in any of our steady-state simulations was $\mu\dot{\gamma}=0.5$ for a state of closely spaced slip planes at $(T, \mu, \dot{\gamma})=(0.01, 0.2, 2.5)$. However, our numerical study was not exhaustive.

A more stringent condition on the parameters applies in the slip-plane regime. In this regime, by definition, no macroscopic fluid regions exist. If $L_{\rm sol}$ is the average length of a solid region within the slip-plane steady-state (i.e.~$L_{\rm sol}$ is the ratio of the system size $N$ to the number of slip planes) then the highest local shear rate at any of the slip planes is 
$(s_j)_{\rm max}\geq\dot{\gamma}L_{\rm sol}$ so that, from Eq.~(\ref{stressStrain}), $\overline{\tau}\geq~\mu\dot{\gamma}L_{\rm sol}$. Hence, the condition $\overline{\tau}\leq~1$ for non-yielding of the solid regions gives 
$\mu\dot{\gamma}\leq~1/L_{\rm sol}$, which is a more stringent condition than that for survival of any micro-solid structures. The definition of the slip-plane regime requires $L_{\rm sol}\geq~2$, the equality occurring in the limit where half of the inter-rotor spaces are slip planes, so that each rotor is solidly bonded to only one neighbour. Hence the condition for the slip-plane phase is $\mu\dot{\gamma}\leq~0.5$. As noted above, this upper bound was realised in one of our simulations (see Fig.~\ref{collapse}), but never exceeded.

\subsection{Linearization}

The full equations of motion (\ref{gapEoM} and \ref{torque}) are non-linear in the unknown functions $\Delta\theta_j(t)$. To find solutions, we shall linearize them. This is not simply a case of replacing the sinusoid in Eq.~\ref{torque} by its small-angle limit, as one might at equilibrium, since, in the boundary-driven system, (some of) the relative angles $\Delta\theta_j$ are ever-increasing functions of time.

At steady state, $\Delta\theta_j(t)$ can be written in terms of a part that grows linearly at rate $s_j$ (the local shear rate), a constant offset $c_j$, and a bounded fluctuating part $\e_j(t)$ with zero mean, thus:
\begin{equation}
\label{timeDependence}
	\Delta\theta_j(t) = s_j \,t + c_j + \e_j(t).
\end{equation}
Then the (still exact) equations of motion become
\begin{subequations}
\begin{equation}
\label{epsilon}
	\ddot{\e}_j(t) = \tau_{j+1} + \tau_{j-1} -2\, \tau_{j}
\vspace{-4mm}
\end{equation}
\begin{equation}
\label{tau}
	\tau_j = \sin\!\big(s_j \,t + c_j + \e_j(t)\big) + \mu s_j +\mu\, \dot{\e}_j(t) + \eta_j(t).
\end{equation}
\end{subequations}
Time averaging Eq.~\ref{tau} and using the boundedness of $\e_j(t)$ allows Eq.~\ref{stressStrain} to be recast as
\begin{equation}
\label{stressRelation}
	\overline{\tau} = \mu s + \left\langle \sin\!\big(s_j \,t + c_j + \e_j(t)\big) \rangle\right.
\end{equation}

Next, we expand $\e_j(t)$ in terms of Fourier modes with coefficients $\hat{\e}_j(\omega)$ thus:
\begin{equation}
\label{Fourier}
	\e_j(t) = \sum_\omega \hat{\e}_j(\omega)\, e^{i\omega t}
\end{equation}
where the summation is over a discrete but infinite set of frequencies $\{\omega\}$ that are not necessarily equally (or even finitely) spaced, since the motion may be aperiodic. A continuous Fourier transform was not used in Eq.~\ref{Fourier} because $\e_j(t)$ has a discrete spectrum of delta functions, excited by the continual rotation of the interacting pairs of rotors. Similarly, the torque is expanded as
\begin{equation}
\label{torqueFourier}
	\tau_j(t) = \sum_\omega \hat{\tau}_j(\omega)\, e^{i\omega t}.
\end{equation}
Real-valued $\e_j(t)$ and $\tau_j(t)$ require the positive and negative-frequency spectral components to respect the symmetry 
\mbox{$\hat\e_j(-\omega) = \hat\e_j^*$} and
\mbox{$\hat\tau_j(-\omega) = \hat\tau_j^*$} where an asterisk ($^*$) indicates the complex-conjugate.

The inverse transformations (Fourier analysis) can be performed by applying temporal averaging on the infinite time domain
\begin{equation}
	\< e^{-i\omega t}\e_j(t) \> 
	= \sum_{\omega'} \hat{\e}_j(\omega') \< e^{i(\omega'-\omega) t} \>
	= \hat{\e}_j(\omega)
\end{equation}
since 
$\< e^{i\omega t} \> = \delta(\omega, 0)$ 
where the Kr\"{o}necker delta has its usual definition,
\[
	\delta(\omega', \omega) = \left\{
	\begin{array}{ll}
		1 & \quad\mbox{ if } \omega' = \omega \\
		0 & \quad\mbox{ otherwise},
	\end{array}
	\right. 
\]
despite the fact that it arguments are non-integer. 
Note that $\hat{\e}_j(0) = 0$ and $\hat{\tau}_j(0) = \overline{\tau}$ and Eq.~\ref{epsilon} becomes
\begin{subequations}
\begin{equation}
\label{epsilonSpectrum}
	-\omega^2\, \hat\e_j(\omega) = \hat\tau_{j+1}(\omega) + \hat\tau_{j-1}(\omega) - 2\hat\tau_j(\omega).
\end{equation}

Analysis in this section is thus far exact. Henceforth, we shall assume, as observed in many of the steady states simulated, that fluctuations about constant motion (with respect to time) are small, and so linearize Eq.~\ref{tau} in $\e_j(t)$, yielding the torque-spectrum
\begin{equation}
\label{tauSpectrum}
\begin{split}
	\hat\tau_j(\omega) & =  \mu\, s_j\, \delta(\omega, 0)
	+ \tfrac{1}{2}i b_j^* \delta(\omega, -s_j) - \tfrac{1}{2}i b_j \delta(\omega, s_j)	
	+ \hat\eta_j(\omega) \\
	& + i\omega\mu\hat\e_j(\omega) + \tfrac{1}{2}b_j \hat\e_j(\omega-s_j)+\tfrac{1}{2}b_j^* \hat\e_j(\omega+s_j)
\end{split}	
\end{equation}
\end{subequations}
where $\hat\eta_j(\omega)\equiv\<\eta_j(t)\,\exp(-i\omega t)\>$ and $b_j\equiv \exp(ic_j)$. The zero-frequency component of the torque-spectrum is the mean torque
\begin{equation}
\label{torqueZero}
	\overline{\tau} = \tfrac{1}{2}i(b_j^*-b_j)\, \delta(s_j,0) + \mu s_j 
	+ \tfrac{1}{2}b_j\hat\e(-s_j) + \tfrac{1}{2}b_j^*\hat\e(s_j).
\end{equation}
The first term on the right-hand side of Eq.~\ref{torqueZero} represents the elastic torque in a solid region of the chain (where the local shear rate $s_j$ vanishes), which depends (via $b_j$) only on the mean angle $c_j$ by which neighbours are twisted. This is the only non-zero term for $s_j=0$, in which case Eq.~\ref{torqueZero} evaluates to $\overline{\tau}=\Im(b_j)=\sin(c_j)$. The second term represents the Newtonian contribution to the viscosity. The non-Newtonian contribution of the final two terms arises only (at linear order) from fluctuations at the same frequency as the local shear rate, which synchronise with the periodic traversing of the potential, thus causing non-uniform occupancy of the potential, due to the rotor lingering for longer at some relative angles than others.

In principal, one can solve Eqs~\ref{epsilonSpectrum} and \ref{tauSpectrum} for $\hat\e_j(\omega)$, then substitute $\hat\e_j(s_j)$ and its complex conjugate $\hat\e_j(-s_j)$ into Eq.~\ref{torqueZero} to find the mean torque for any given configuration of local shear rates $\{s_j\}$ and offsets $\{c_j\}$, yielding a spatial constitutive relation for the XY model. 
The uniformity (independence of $j$) of the torque $\overline{\tau}$ then restricts the allowed configurations of $\{s_j\}$ and $\{c_j\}$ to those exhibited by the steady-state phases.

In practice, solution of Eqs~\ref{epsilonSpectrum} and \ref{tauSpectrum} remains difficult in general, and requires some further approximation. We present two alternative methods in sections \ref{EMT} and \ref{runningPair} below. However, for the particular case of a solid region of the chain (one for which $s_j=0$), the Fourier modes become decoupled, so Eq.~\ref{tauSpectrum} simplifies to
\begin{equation}
	\label{solid}
	\hat\tau_j(\omega) = \overline\tau + \hat\eta_j(\omega) 
	+ \left( i\omega\mu + \sqrt{1-\overline{\tau}^2}
	\right) \hat\e_j(\omega)  \mbox{ for } s_j=0.
\end{equation}
In this case, Eq.~\ref{epsilonSpectrum} becomes a discretized noisy wave equation. At zero temperature, is has a simple travelling-wave solution of the form
\begin{equation}
\label{travellingWave}
	\hat\e_{j+1} = \alpha(\omega) e^{\pm i\omega/v(\omega)} \hat\e_j
\end{equation}
with a frequence-dependent (i.e.~dispersive) wave speed $v(\omega)$ due to dissipation and discretization. In the low-frequency, low-dissipation limit, the wave speed varies with overall torque as $v\to(1-\overline{\tau}^2)^{1/4}$, vanishing in the limit of stability $|\overline\tau|=1$. Closed-form expressions for the real values $v$ and $\alpha$ are straightforwardly obtained by substitution of Eqs.~\ref{solid} and \ref{travellingWave} into \ref{epsilonSpectrum}, but are not reproduced here, as they are large. We shall use this travelling-wave solution for the elastic solid in section \ref{runningPair} to find an approximate solution for slip planes. Firstly, we shall develop an approximation for a general steady state.

\subsubsection{Effective Medium Theory}
\label{EMT}

Here, we shall develop a mean-field theory, in which the fluctuations of neighbours $(j+1)$ and $(j-1)$ of a given site $j$ are uncorrelated with it. In Eq.~\ref{epsilonSpectrum} we average (with respect to complex phase) over the fluctuations of the neighbours, so that only their zero-frequency contributions remain, yielding
\begin{equation}
\label{approx}
	\omega^2\, \hat\e_j(\omega) = 2\hat\tau_j(\omega) - 2\overline{\tau}\,\delta(\omega,0).
\end{equation}
This is equivalent to replacing $\tau_{j\pm1}$ by $\overline{\tau}$ in Eq.~\ref{gapEoM}, by placing a single rotor-pair in a medium of constant torque. In a fluid region, where $s_j\neq0$, the one remaining phase constant $c_j$ becomes arbitrary, so we may choose $c_j=0$ implying $b_j=1$. On the other hand, if $s_j=0$, we must retain $c_j$ as a parameter.

Now that we are considering only a single shear rate $s = s_j$, Eq.~\ref{tauSpectrum} and \ref{approx} will generate harmonic spectra, i.e.~only at integer multiples of $s$. So, to simplify the notation, for integer $n$, let us define new coefficients $a_n\equiv\hat\e_j(ns)$ and $\xi_n\equiv\hat\eta_j(ns)$. Hence $a_0=\xi_0=0$ and $a_{-n}=a_n^*$.

So Eq.~\ref{torqueZero} yields
\begin{equation}
\label{inBetween2}
	\overline{\tau} = \sin(c_j)\,\delta(s,0) + \mu s + \Re(a_1)
\end{equation}
where $a_1\to 0$ if $s=0$. For non-zero $s_j$, Eqs.~\ref{approx} and \ref{tauSpectrum} yield
\begin{equation}
\label{inBetween}
\begin{split}
	ns(ns-2i\mu)a_n &=  2(\mu s-\overline{\tau})\delta(n,0) + i\delta(n,-1) - i \delta(n,1) \\
		& + 2\xi_n+a_{n-1}+a_{n+1}.
\end{split}
\end{equation}  

\begin{figure}
  \includegraphics[width = 6.5cm]{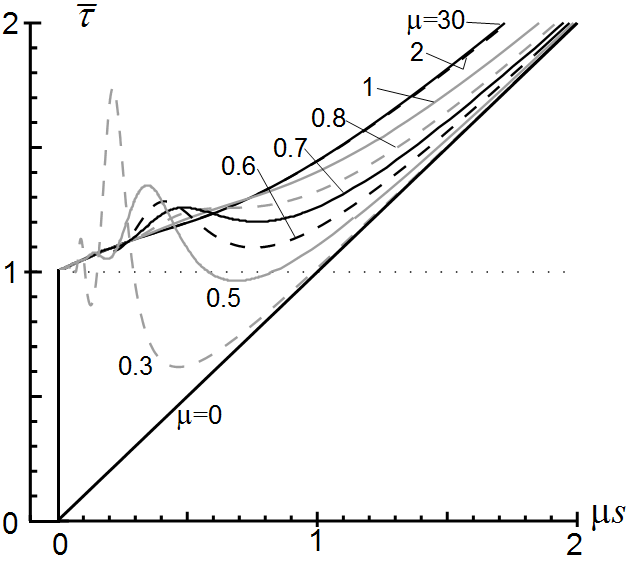}
  \caption{\label{EMTCurves}Torque $\overline{\tau}$ as a function of $\mu s$, the product of friction coefficient and local shear rate, for various (labelled) values of $\mu$ and zero temperature, as given by Eq.~\ref{result} from the approximations of effective medium (mean-field) theory (Eq.~\ref{approx}) and linearization in temporal fluctuations (Eq.~\ref{tauSpectrum}). To distinguish curves at different values of $\mu$, they alternate between continuous and dashed lines, and in pairs of alternating black and grey. To generate the graphs, the first 100 levels of the continued fraction in Eq.~\ref{result} were evaluated (by setting $R_{100}=0$ and iterating Eq.~\ref{recurrence}). Evaluation of further levels was not found to alter the curves perceptibly. Also, higher values of $\mu$ yield curves indistinguishable from the $\mu=30$ curve.}
\end{figure}

To make further progress, in the mean-field spirit, we drop (average over) the noise term $\xi_n$, equivalent to setting zero temperature. For $n=1$, Eq.~\ref{inBetween} yields
\begin{equation}
\label{inBetween3}
	a_1 = \frac{i}{\frac{a_2}{a_1} - s(s-2i\mu)}.
\end{equation}
For $n\geq2$, Eq.~\ref{inBetween} yields a simple recurrence relation for the ratio of successive harmonic amplitudes $R_n\equiv~a_n/a_{n-1}$ for $n>1$,
\begin{equation}
\label{recurrence}
	\frac{1}{R_n} = ns(ns-2i\mu)-R_{n+1}.
\end{equation}
Using Eq.~\ref{recurrence} to define the as-yet undefined variable $R_1$ allows us to combine Eqs.~\ref{inBetween2}, \ref{inBetween3} and \ref{recurrence} into a neat expression of the fluid's consititutive relation, in terms of the imaginary part of a continued fraction,
\begin{eqnarray}
	\overline{\tau} &=& \sin(c_j)\,\delta(s,0) + \mu s +\Im(R_1) 	\nonumber  \\
\label{result}
	R_1 &=& \frac{1}{s(s-2i\mu)-\frac{1}{2s(2s-2i\mu)-\frac{1}{3s(3s-2i\mu)-\frac{1}{\ldots}}}}.
\end{eqnarray}
Equation~\ref{result} is plotted in Fig.~\ref{EMTCurves} for various values of the friction coefficient $\mu$.

Note that we find values of mean torque $\overline{\tau}$ that depend on $\mu$, not only on the combination \mbox{$\mu s$} and $T$ as approximated in appendix \ref{scaling}. This is not unexpected, since the scaling approximation in the appendix requires the correlated motion of several nearby rotors to behave as a continuum, whereas the mean-field theory yielding Eq.~\ref{result} analyses only a single rotor-pair. It is worth remarking that the scaling collapse of Fig.~\ref{collapse} is only approximate, so that the qualitative effect of varying $\mu$ in Fig.~\ref{EMTCurves} is not inconsistent.

A local constitutive relation, or ``flow curve" (a fluid's stress as a function of {\em local} shear rate) of the kind plotted in Fig.~\ref{result} has a standard interpretation \cite{Olmsted}. A negative gradient of the curve indicates an unstable regime. No fluid can remain uniform at a shear rate where the gradient is negative, as the unbalanced stresses on a region of the fluid that is infinitesimally perturbed from uniformity will act to enhance the perturbation. Furthermore, because the time-averaged torque must be uniform at steady state, any inhomogeneous fluid must exhibit a coexistence of local shear rates at equal values of $\overline{\tau}$. Hence, a shear-banding state, or a solid-fluid coexistence, can only exist at values of the parameters $\mu$ and $s$ for which a horizontal line has multiple intersections with the local flow curve. 

The above criteria are not sufficient to uniquely define the extent of the coexistence behaviour from the curves in Fig.~\ref{EMTCurves}, but do put bounds on it. Specifically, since the overall shear rate $\dot\gamma$ is a volume-weighted average of the coexisting local shear rates, a uniform steady state cannot exist at an overall shear rate where the flow curve has negative slope. And, inverting the flow curve, a uniform steady state must exist wherever the local shear rate is single-valued as a function of $\overline\tau$. 

For $s=0$, Eq.~\ref{result} describes a solid region capable of supporting any torque in the range $|\overline\tau|<1$ (depending on the value of $c_j$), corresponding to the vertical line segment from $(0,0)$ to $(0,1)$ in Fig.~\ref{EMTCurves}.
The horizontal dotted line is the upper limit $\overline\tau=1$ on the solid regions' torque. So curves that cross the dotted line represent possible solid-fluid coexistences.

Hence, Fig.~\ref{EMTCurves} predicts (for $T=0$) only uniform states for $\mu\geq0.8$. On decreasing $\mu$, a critical point is predicted just below $\mu=0.8$ and $\mu\dot\gamma\approx0.7$, where a negative gradient first appears. Below this critical value of $\mu$, shear-banding takes place for a finite range of shear rates. Only at lower values (below $\mu\approx0.53$) is solid-fluid coexistence possible, according to the approximate solution.

Comparing with the simulation results, and extrapolating the simulated trends to zero temperature where simulation results are difficult to obtain due to long-lived transients, we see that the approximate theory, captures some qualitative features. In particular, in simulations, a uniform state exists at high $\mu$ and, on decreasing $\mu$ the shear-banded state is entered via a continuous transition, whereas the ``freezing" transitions (to solid-fluid coexistence or slip planes) appear to be first-order, and occur at lower $\mu$. 

\begin{figure}
  (a) \\ 
  \includegraphics[width = 8cm]{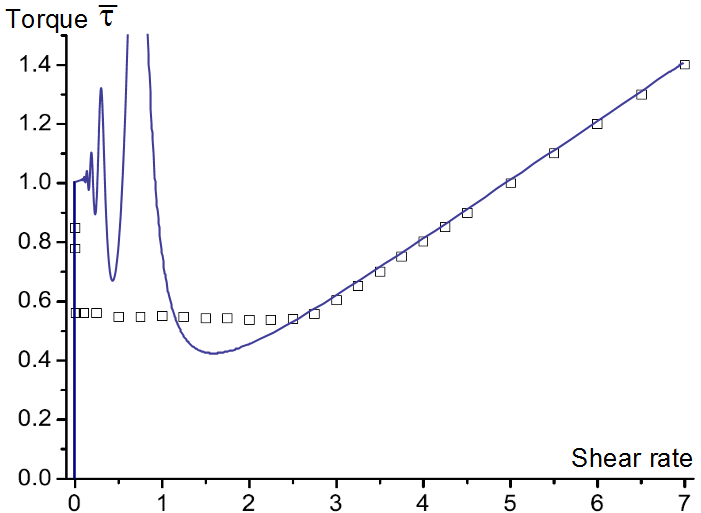} \\
  (b) \\ 
  \includegraphics[width = 7.8cm]{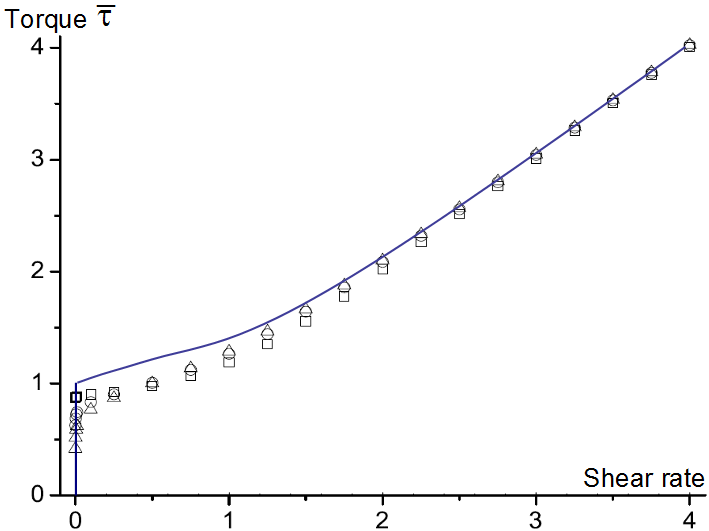}
  \caption{\label{compareData} Comparison of mean torque between theory and simulation. Continuous curves: the mean-field predictions as a function of {\em local} shear rate, with thermal fluctuations neglected. Symbols: the simulation data of Fig.~\protect\ref{stress-strain}, plotted against {\em global} shear rate, with squares representing the lowest temperature ($T=0.01$). Theory and simulations are both evaluated for (a) $\mu=0.2$ and (b) $\mu=1.0$.}
\end{figure}

Furthermore, quantitative comparison can be made with values of the mean torque measured in simulation. For the torques shown in Fig.~\ref{stress-strain}a, from simulations at $\mu=0.2$, temperature has a significant effect. Only the lowest-temperature data (shown by squares) should be compared with the mean-field theory, which is effectively a zero-temperature theory. In agreement with the theoretical curve, the data (Fig.~\ref{compareData}a) are almost indistinguishable from a straight line (a Newtonian fluid) for all high shear rates. 

A phase transition to a coexisting state is characterized, in the simulation data, by a transition to a region of vanishing gradient. Note that the data are plotted against the {\em global} shear rate, averaged over the coexisting regions, whereas the mean-field curve is a function of {\em local} shear rate, and so can only put bounds on the volume-weighted mean of coexisting phases. (Recall that coexistence is necessary wherever the local constitutive relation has a negative gradient, and forbidden wherever local shear rate is single-valued as a function of $\overline\tau$.) Hence, Fig.~\ref{stress-strain}a demonstrates that the mean-field theory is completely consistent with the measured phase transition(s) and quantitatively fairly accurate at $\mu=0.2$.

For any single-phase regime, the theory makes a unique prediction. A single phase exists for the data in Fig.~\ref{stress-strain}b at $\mu=1.0$, for which temperature has little effect. They are compared with the mean-field curve in Fig.~\ref{compareData}b. Although imperfect, the mean-field theory is not far from the data.

\subsubsection{An isolated slip plane}
\label{runningPair}

We can solve for the motion of the fully interacting linearized system (i.e.~no longer neglecting the correlated fluctuations $\e_j$ and offsets $c_j$ of neighbouring rotor-pairs) for the specific case of the slip-plane phase at $T=0$, subject to the approximation, to be introduced below, that the motion is dominated by a single frequency (the rotation rate of the local slip plane). We consider an isolated running pair (where $\Delta\theta_j$ increases with time as Eq.~\ref{timeDependence} with finite $s_j$) in a chain of otherwise locked ($s_j=0$) rotors. Let us assume that the slip planes are sufficiently far apart that we may consider only one isolated slip plane. 

The running pair (slip plane), which we locate at $j=0$ without loss of generality, sends out small-amplitude waves into the surrounding solid chain of rotors. The motion of the locked rotor pairs, for all $j\geq~2$ is given in terms of $\hat\e_1$ by Eq.~\ref{travellingWave}, and similarly the variables at negative $j$ are eliminated in favour of $\hat\e_{-1}$. Finally, $\hat\e_{\pm1}$ are eliminated in favour of $\hat\e_0$ by applying Eq.~\ref{epsilonSpectrum} for $j=\pm1$ with the torque given by Eq.~\ref{solid} for $j\neq0$ and Eq.~\ref{tauSpectrum} for $j=0$. 

Given that the pair at $j=0$ is the only running pair, the absolute phase of its motion is irrelevant, so we may set $c_0=0$ hence $b_0=1$, so that Eq.~\ref{tauSpectrum} for the torque at $j=0$ simplifies a little. Nevertheless it couples Fourier coefficients at all harmonic frequencies, thus eventually yielding a recurrence relation for these harmonics, analogous to the one found for effective-medium theory (Eq.~\ref{inBetween}), but much less neat (hence we do not show it here). Unfortunately, we have not found a full solution to this recurrence relation, as we did in section \ref{EMT}. So we further approximate that the motion is dominated by the frequency $s$, and drop all higher harmonics (setting $\hat\e_0(ns)=0$ for $n\neq\pm1$). 

\begin{figure}
  (a) \\ \includegraphics[width = 5.7cm]{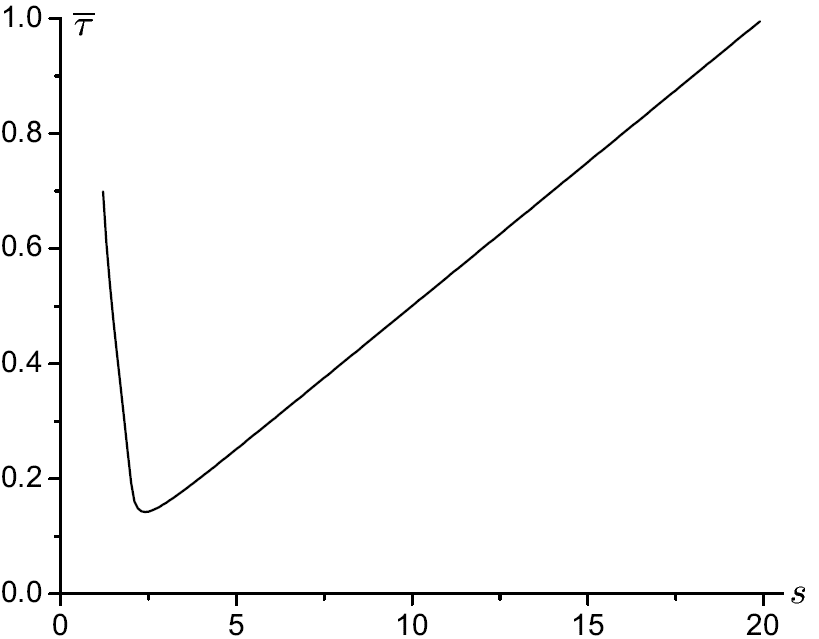} \\
  (b) \\ \includegraphics[width = 5.7cm]{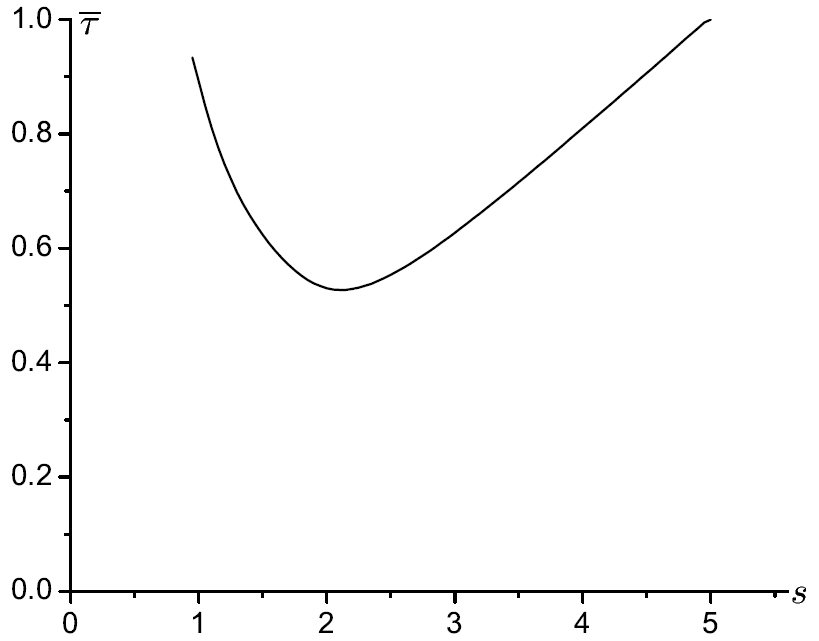} \\
  (c) \\ \includegraphics[width = 6.2cm]{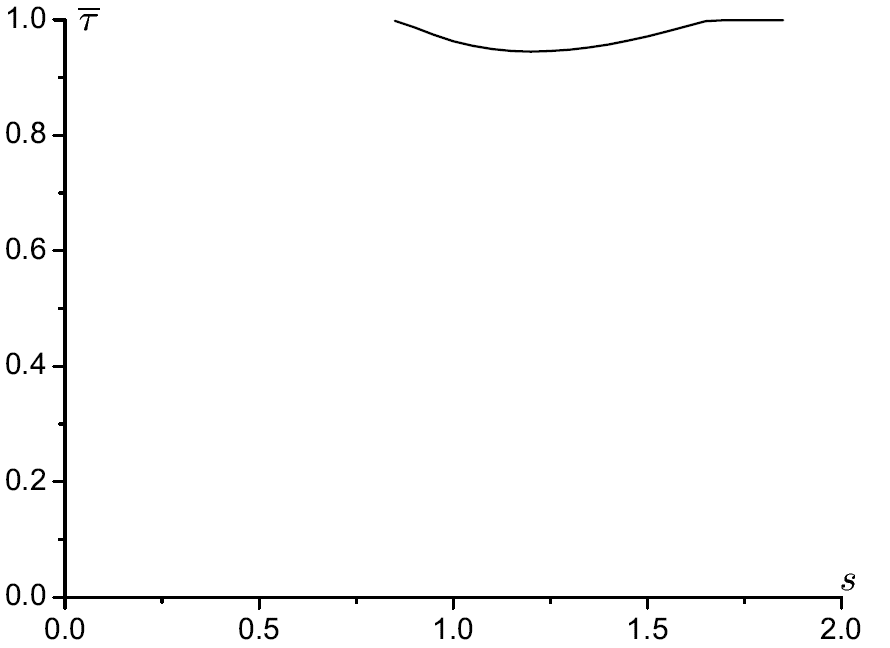}
  \caption{\label{runningConstitutive}The mean torque in a system consisting of a single isolated running rotor pair. The solution only exists at torques $|\overline\tau|\leq1$, and thus extends over a finite domain in $s$, that depends on the friction coefficient $\mu$. (a) $\mu=0.05$, (b) $\mu=0.2$, (c) $\mu=0.5$.}
\end{figure}

The resulting relationship between mean torque $\overline\tau$ and rotation rate $s$ of the running pair is plotted in Fig.~\ref{runningConstitutive} for various $\mu$.

Recall (as in section \ref{bounds}) that, if the (effectively isolated) slip planes are separated by an average distance of $L_{\rm sol}$ rotors, then the overall shear rate is related to the local shear rate at the slip plane by $\dot\gamma=s/L_{\rm sol}$. Since $s$ is confined to a finite (and, in some cases small) range of values (see Fig~\ref{runningConstitutive}) --- indeed only to the subset of stable values for which the gradient of $\overline\tau(s)$ is positive --- while $\dot\gamma$ is fixed by the boundary conditions, it follows that $L_{\rm sol}$ is confined to vary approximately with $\dot\gamma^{-1}$. Hence the number of slip planes varies approximately linearly with $\dot\gamma$, as observed in the simulation results of Fig~\ref{slipPlaneCount}. The gap (in $s$ values), between the stable regimes at zero shear rate and at finite $s$, is responsible for the existence of the slip-plane phase, which is unable to re-distribute its localized points of high shear rate into more diffuse regions of lower shear rate.

While different in detail from the curves in Fig.~\ref{EMTCurves}, we note that the solutions in Fig.~\ref{runningConstitutive} are close in value to those in Fig.~\ref{EMTCurves} which were derived (at mean-field level) for local shear rate $s$ in an arbitrary phase. This agreement therefore lends further support to the validity of the earlier mean-field result.

\begin{figure}
  (a) \\ \includegraphics[width = 6.5cm]{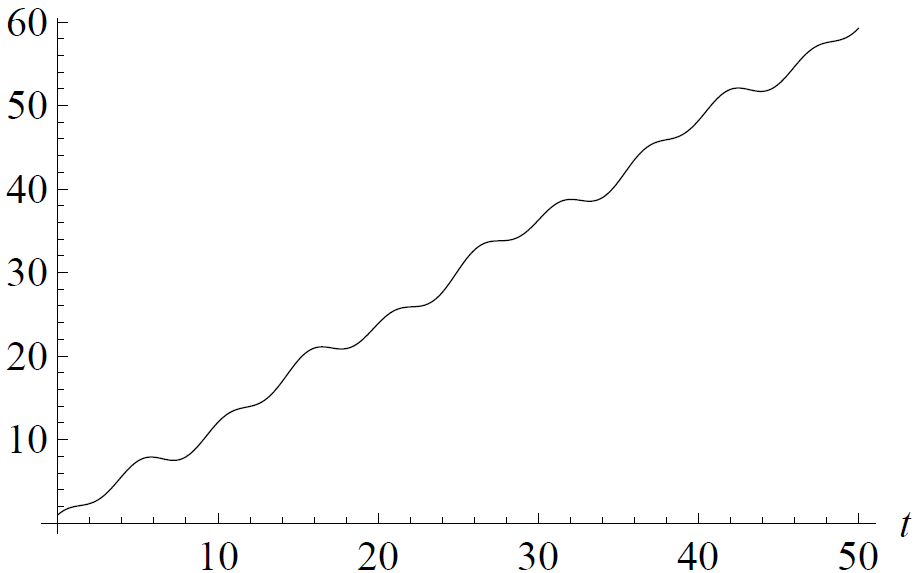} \\
  (b) \\ \includegraphics[width = 7cm]{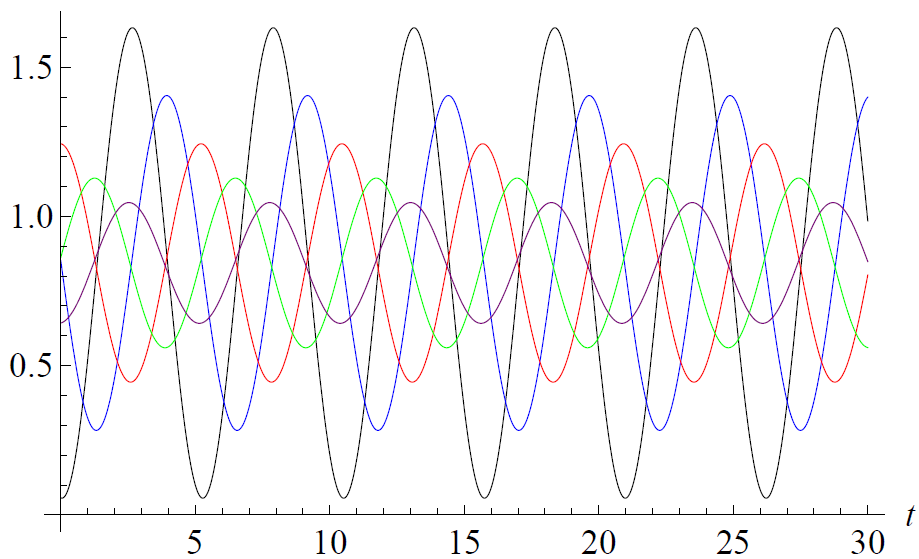}
  \caption{\label{wiggles}The relative angle of (a) the running pair of rotors and (b) the first five rotor pairs to the right of the running pair as functions of time, for $s = 1.2$
  and $\mu = 0.2$.}
\end{figure}

Example trajectories for the running pair of rotors and some of its neighbours in the solid region are shown in Fig.~\ref{wiggles}. The further into the solid region, the smaller the amplitude of the fluctuations about the average value. The fluctuations are successively out of phase by the same amount, indicating a travelling torsional wave, like those visible in the simulation data of Fig.~\ref{space-time-SLIP} as diagonal lines emanating a short distance from the slip plane. Hence, we see that each slip plane disturbs its environment
via small-amplitude waves that decay as they propagate. Two nearby slip planes will interact appreciably only if their separation is small compared with the decay length of the torsional waves.

\section{Conclusions and prospects for further work}
\label{conclusions}

It is remarkable how closely the shear flow of the angular-momentum-conserving 1D XY model mimics the rich behaviour of non-Newtonian complex fluids. We have observed that, independent of its initial conditions, the three-parameter model exhibits reproducible steady-states with four main types of velocity profile, as shown in Fig.~\ref{examples}: uniform flow, shear-banding, solid-fluid coexistence and slip-plane states. There may be yet other, more exotic states in small regions of parameter-space.

All four of the model's main flow regimes are routinely seen in diverse types of soft matter, which have hitherto each been analysed using different {\em ad hoc} phenomenological models that are often quite complex. 

The macroscopic phenomenology of non-Newtonian fluids is typically captured theoretically using constitutive
models, normally tensorial models that define the stress tensor in terms of the
mesoscopic structure, for example the (diffusive) Johnson-Segalman model \cite{Johnson77,Olmsted00}.
The dynamics of the structure is specified and then the model can be solved for
the flow properties either in or out of the steady state, usually numerically.
While tensorial models are needed to describe real fluids, simpler non-tensorial
models have previously been defined phenomenologically, without microscopic degrees of freedom (e.g.~\cite{Fielding04,Dubbeldam09}), and can
reproduce phenomena such as the instability in the constitutive curve described above.

The reproduction of so many non-Newtonian macroscopic phenomena in a single, simple, microscopic model was quite unexpected, and without precedent. It offers further insights into the microscopic and macroscopic features of flow-induced phase transitions.

By numerically charting some of the phase diagram and explaining some of the behaviour by linearization and mean-field analysis, we have demonstrated the XY model's great richness and its usefulness as an idealized model of non-equilibrium fluids. Furthermore, we have shown that there is much scope for further research on this system, which we would encourage others to investigate.

In particular, more numerical work is required, to explore the phase diagram more comprehensively and in greater detail, and also to fully characterize the nature of the various phase transitions, which will contribute to a wider understanding of non-equilibrium condensed matter in general. Once those phase transitions have been characterized, it will be interesting to make comparative experimental studies of complex fluids that exhibit equivalent behaviour.

Theoretical progress is possible in a number of directions. It may be feasible to solve the linearized version of the model (i.e.~to first order in deviations $\varepsilon_j$ from constant but non-uniform shear rate $s_j$) without applying the effective medium approximation that removes spatial features. Alternatively, a perturbative expansion in $\varepsilon_j$, perhaps to second or third order, would reveal localized interactions that govern the curvature of velocity profiles such as Fig.~\ref{examples}b and create an effective interfacial energy between coexisting phases, leading to their macroscopic separation. 

It would be appealing to find an approximate continuum version of the model, expressed in terms of differentiable fields that are functions of continuous position $x$ and time $t$ only. The difficulty is that, although the local torque $\tau(x,t)$ can, in some regimes, be assumed smoothly varying (as in Appendix \ref{scaling}), the relative angle $\Delta\theta_j$ cannot, and must remain indexed by a discrete index $j$ (although its time-derivative may be made continuous). The reason is that $\Delta\theta_j$ is periodic modulo $2\pi$, so that the torque (which depends on $\Delta\theta_j$) is invariant under a full turn of one rotor relative to its neighbour. A continuous field $\Delta\theta(x)$ would acquire a topological defect with increasing winding number each time a full turn was introduced, which would be appropriate for modelling an elastic solid with memory of total acquired strain (i.e.~a twisting rubber belt), but not a fluid under continuous shear flow. Overcoming this difficulty, to derive a Ginsburg-Landau-like model for the steady-state profile of local shear rate $s$ would require approximately ``integrating out" (i.e.~solving for) the fluctuations $\varepsilon_j$.

Some other theoretical approaches with potential for further progress, including some ideas for renormalization group analysis, can be found in Ref.~\cite{WelshThesis}, along with some further numerical data on this elegant model.

\section{Acknowledgements}

We are grateful for the advice of Adrian Baule. Parts of this work were funded by a Royal Society University Research Fellowship, an EPSRC grant GR/T24593/01
and an EPSRC doctoral training grant.

\medskip
\appendix 
\begin{center}
{\bf APPENDIX}
\end{center}
\vspace{-1cm}
\subsection{\label{scaling}Approximate scaling}

To investigate effects of re-scaling the model's parameters, let us non-dimensionalize time by measuring it in units of reciprocal shear rate. We define the reduced-time $\hat{t}\equiv\dot{\gamma}t$, and note that the noise function $\eta_j(t)$ scales non-trivially with time and with friction coefficient and temperature. Writing it explicitly as a function, $\eta_j(t,\mu,T)$, we note, from Eq.~\ref{noise} the following invariance with respect to re-scaling of its arguments:
\begin{equation}
	\eta_j(at,b\mu,\tfrac{a}{b}T) \leftrightarrow \eta_j(t,\mu,T)
\end{equation}
where the equivalence $\leftrightarrow$ means that statistical properties (specifically, all moments) of the functions are equal. Hence we may write
\begin{equation}
	\eta_j(t,\mu,T) = \eta_j(\tfrac{\hat{t}}{\dot{\gamma}},\mu,T)
	\leftrightarrow \eta_j(\hat{t},\mu\dot{\gamma},T).
\end{equation}
So Eq.~\ref{torque} becomes
\begin{equation}
\label{rescaledtorque}
    \tau_j = \sin \Delta\theta_j  + \left(\mu\dot{\gamma}\right) \frac{\td\Delta\theta_j}{\td\hat{t}} 
    + \eta_j(\hat{t},\mu\dot{\gamma},T).
\end{equation}
So we see that the formula for the torque retains the form of Eq.~\ref{torque} under the temporal re-scaling $t\to\hat{t}=\dot{\gamma}t$ if the system parameters are re-scaled according to
\begin{equation}
	(T,\mu,\dot{\gamma}) \to (\hat{T},\hat{\mu},\hat{\dot{\gamma}}) = (T,\mu\dot{\gamma},1)
\end{equation}
where the transformation of $\dot{\gamma}$ follows from the fact that it is the positional average of $\Delta\dot{\theta}_j$. 

Although Eq.~\ref{torque} is invariant under the above transformation, Eq.~\ref{gapEoM} is not. To find a symmetry of this pair of equations, we must first approximate Eq.~\ref{gapEoM}, by appealing to the fact that its right-hand side is a discrete version of a second derivative with respect to position. If we assume that torque varies smoothly with position, then we may replace the discrete index $j$ by a continuous position $x$ and approximate the right-hand side of Eq.~\ref{gapEoM} by a second derivative.
Under that assumption, if we define our unit of length to be the distance between rotors, then Eq.~\ref{gapEoM} becomes
\begin{equation}
\label{continuum}
  \frac{\td^2\Delta\theta_j}{\td t^2} \approx \frac{\partial^2 \tau(x,t)}{\partial x^2}.
\end{equation}
Our observations of the steady-state dynamics lead us to believe that the assumption of smoothly-varying torque is a reasonably realistic approximation in some but not all regimes. 

As explained in section~\ref{conclusions}, although $\tau(x,t)$ can (in some regimes) be assumed smoothly varying, the relative angle $\Delta\theta_j$ cannot.
Hence the left-hand side of Eq.~\ref{continuum} remains a function of $t$, indexed by $j$, and the right-hand side is evaluated at $x=j$.

Under the temporal re-scaling $t\to\hat{t}=\dot{\gamma}t$, Eq.~\ref{continuum} transforms to
\begin{equation}
	\frac{\td^2\Delta\theta_j}{\td\hat{t}^2}
	=\frac{1}{\dot{\gamma}^2}\frac{\td^2\Delta\theta_j}{\td~t^2}
	\approx \frac{1}{\dot{\gamma}^2} \frac{\partial^2 \tau(x,t)}{\partial x^2}
\end{equation}
which can be written as
\begin{equation}
\label{rescaledcontinuum}
  \frac{\td^2\Delta\theta_j}{\td\hat{t}^2} \approx \frac{\partial^2 \tau(x,t)}{\partial \hat{x}^2}
\end{equation}
if we define a re-scaled position $\hat{x}\equiv\dot{\gamma}x$.

Finally, then comparing Eqs.~\ref{rescaledcontinuum} and \ref{rescaledtorque} with \ref{continuum} and \ref{torque}, we observe that, for states with smoothly varying torque, a system with parameters	$(T,\mu,\dot{\gamma})$  respects approximately the same equations of motion as a system with parameters $(T,\mu\dot{\gamma},1)$ and re-scaled length and time. Hence, these systems will exhibit the same steady-state phase behaviour.
Therefore, we expect some partial data-collapse when plotting the phase diagram on axes of $T$ and $(\mu\dot{\gamma})$ only. As seen in Fig.~\ref{collapse}, the resulting data-collapse is not complete, indicating that local details of the torque variation control a significant amount of the macroscopic behaviour.
Nevertheless, the scaling analysis in this appendix --- particularly the exact scaling in Eq.~\ref{rescaledtorque} --- may prove useful for future studies.

\end{document}